# Single-photon distributed free-space spectroscopy

SAIFEN YU,[1,†] ZHEN ZHANG,[1,†] HAIYUN XIA,[1,2,*] XIANKANG DOU,[1,2,*] MANYI LI,[1] TIANWEN WEI,[1] LU WANG,[1] PU JIANG,[1] YUNBIN WU,[1] CHENGJUN ZHANG,[3] LIXING YOU,[3] YIHUA HU,[4] TENGFEI WU,[5] LIJIE ZHAO,[1] MINGJIA SHANGGUAN,[1] LEIGANG TAO,[2] AND JIAWEI QIU[1]

[1] *School of Earth and Space Science, University of Science and Technology of China, Hefei 230026, China*
[2] *Hefei National Laboratory for Physical Sciences at the Microscale, Heifei 230026, China*
[3] *Shanghai Institute of Microsystem and Information Technology, Chinese Academy of Sciences, Shanghai 200050, China*
[4] *State Key Laboratory of Pulsed Power Laser Technology, National University of Defense Technology, Hefei 230037, China*
[5] *Changcheng Institute of Metrology & Measurement, Aviation Industry Corporation of China, Beijing 100095, China*
[†]*These authors contributed equally to this work*
*\*Corresponding author: hsia@ustc.edu.cn; dou@ustc.edu.cn*

**Spectroscopy is a well-established nonintrusive tool that has played an important role in identifying substances and quantifying their compositions, from quantum descriptions to chemical and biomedical diagnostics. Challenges exist in accurate measurements in dynamic environments, especially for understanding chemical reactions in arbitrary free-space. We develop a distributed free-space spectroscopy realized by a comb-referenced frequency-scanning single-photon lidar, providing multidimensional (time-range-spectrum) remote sensing. A continuous field experiment over 72 hours is deployed to obtain the spectra of multiple molecules ($CO_2$ and HDO) in free-space over 6 km, with a range resolution of 60 m and a time resolution of 10 min over a spectrum span of 30 GHz. The $CO_2$ and HDO concentrations are retrieved from the spectra acquired. This distributed free-space spectroscopy holds much promise for increasing knowledge of atmospheric environments and chemistry research, especially for complex molecular spectra evolution in any location over large areas.**

Molecular spectroscopy dates back to Newton's division of sunlight into seven colors from red to purple by a prism. With the advent of quantum mechanics and lasers, high-resolution spectroscopy has developed rapidly in many fundamental domains, especially from the quantum description of matter to nonintrusive diagnostics of various media [1]. Thanks to the molecular fingerprints contained in characteristic spectra [2], Fourier transform spectroscopy, differential absorption spectroscopy, intracavity laser absorption spectroscopy and cavity ring-down spectroscopy identify and quantify substances on the basis of ultrasensitive spectroscopic measurements ranging from chemical and biomedical diagnostics to atmospheric sensing [3-6]. Additionally, spectrometers are widely used as payloads on low earth orbit satellites for global monitoring of atmospheric trace gases [7-8]. However, the approaches mentioned above provide either local or column continuous measurements or instantaneous "snapshot" regional sampling measurements [9,10]. For multiple molecule identification, trace gas exchange characterization and atmospheric chemical reactions that vary episodically and spatially, distributed free-space spectra in continuous observations over large regions, especially inaccessible locations, should be obtained.

Light detection and range (lidar) methods actively emit light pulses and gives a unique possibility to remotely sense three-dimensional distributions of molecules. Differential absorption lidar (DIAL) emits two high-energy laser pulses at slightly different frequencies: one is strongly absorbed by the molecule under investigation, and the other is weakly absorbed [11]. Multifrequency DIALs built on this basis effectively obtain broadband absorption spectra for molecular measurements. Examples include multiple fixed-wavelength distributed-feedback laser diodes (DFB-LDs) [12], scanning DFB monitored by heterodyne detection [13], injection-seeding optical parametric oscillators (OPOs) [14] and nest dual-cavity doubly resonant optical parametric oscillators (NesCOPOs) with multifrequency output [15]. Usually, spectra integrated over the optical path are obtained, which sacrifices range resolution and ignores local variations. A DIAL system is optimized for specific substance detection and is restricted by the requirement of a high-power laser at the absorbing optical frequency.

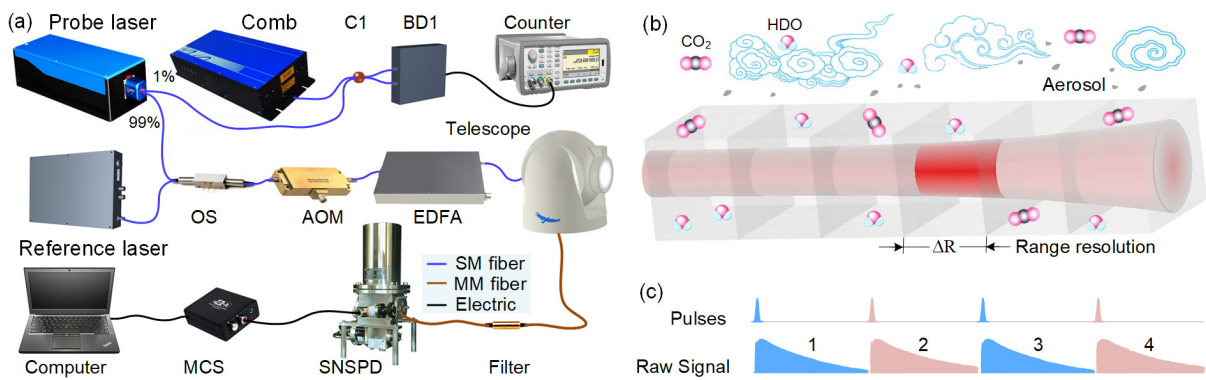

**Fig. 1.** Optical layout. (a) Experimental set-up. The probe laser, together with a reference laser, is switched and gated by OS and AOM to ensure real-time atmospheric correction. The optical frequency comb is used to generate beat notes with the probe laser. Therefore, the probe laser can be programmatically tuned and locked. (b) Lights propagating in the atmosphere. The path length in the red sections represents the range resolution of ΔR, and the spectra within this range in the whole free-space can be obtained. (c) Time sequence of the time-division multiplexing technique.

Recently, stabilized optical frequency combs have revolutionized time and frequency metrology, opening up novel opportunities for precise spectroscopy with resolved individual comb lines and a broad spectral range from ultraviolet to infrared [16]. For instance, Fourier transform spectroscopy with frequency combs [3], cavity-enhanced dual-comb spectroscopy [6], frequency comb-assisted diode laser comb spectroscopy [17], dual-comb spectroscopy [18-21], and phase-locked [22] and injection-locked [23] optical synthesizers based on combs have been used. Such comb-assisted spectrometers have realized broadband and precise molecular spectra measurements on site or regional.

We demonstrate a frequency-scanning single-photon lidar for realizing distributed free-space spectroscopy (DFSS) with multiple dimensions. First, a comb-referenced technique is used for scanning and locking the frequency of an external cavity diode laser (ECDL). The number of scanning steps, the frequency intervals and the span can be redefined to adapt to the spectra of substances under investigation. Second, a superconducting nanowire single-photon detector (SNSPD) [24,25] with high light detection sensitivity and low dark noise count is adopted to detect weak backscattering signals from the atmosphere. To suppress the effect of turbulence on the free-space to fiber coupling efficiency of the telescope, a large active area SNSPD is manufactured with a diameter of 50 μm. Thanks to the high signal-to-noise ratio of SNSPDs, we can implement DFSS over a wide range of optical frequencies using a low-power fiber laser. Third, considering the variation in atmospheric conditions and instability of the system during spectrum scanning, a reference laser and a probe laser are emitted alternately by employing the time-division multiplexing (TDM) technique. Finally, DFSS can analyse multiple substances in multiple dimensions, offering the ability to remotely sense consecutive spectra at any point in free-space, especially somewhere inaccessible. DFSS naturally holds much potential for improving human knowledge in greenhouse gas monitoring, leakage warning and atmospheric chemistry reaction research.

Considering the importance of trace gases in atmospheric transport [26], we take $CO_2$ as representative substance due to its strong greenhouse effect. The $CO_2$ R16 line at 190.667 THz is chosen with a span of 30 GHz, containing two weak HDO lines (Fig. S1(b)). A diagrammatic view of DFSS is shown in Fig. 1. A single frequency tunable ECDL covering 185.185 ~ 196.078 THz is used as a probe laser. One percent of the probe laser is combined with an optical frequency comb for heterodyne detection, enabling the frequency of the probe laser to be precisely tuned and locked by the program (Supplement 1, Fig.S1). The other 99% of the probe laser and a reference laser are chosen alternatively by a fast optical switch (OS) and chopped into laser pulses through an acoustic-optical modulator (AOM) to give a pulse repetition rate of 20 kHz and a full width at half maximum (FWHM) of 400 ns. Such an optical layout guarantees that the probe pulse and the reference pulse have exactly the same shape. The time sequence of TDM is shown in Fig. 1(c), where the probe pulse is tagged in odd numbers (1, 3, ...) and the reference pulse is tagged in even numbers (2, 4, ...). The pulse energies of chopped pulses are amplified by an erbium-doped fiber amplifier (EDFA) to 40 μJ. Then, the laser pulses are pointed to the atmosphere through a collimator with a diameter of 80 mm. As the probe laser scans the $CO_2$ and HDO absorption spectra, the reference laser remains at a stable frequency of 190.652 THz. After that, the atmospheric backscattering signal is received and coupled into an optical fiber by using a telescope with a diameter of 256 mm. Finally, the signal is fed to the SNSPD. The SNSPD is divided into 9 pixels, and each pixel consists of two superconducting nanowire avalanche photodetectors, which improves the maximum count rate of the detector. The total system detection efficiency of the SNSPD array is 31.5% at 190.667 THz, and the dark count rate is approximately 100 Hz.

As shown in Fig. 1(a), the probe and reference pulses share the same optical path, SNSPD and electric circuits, resulting in only the optical frequency differing for spectrum analysis. Eventually, as shown in Fig. 1(b), the absorption spectra of gases under investigation are obtained by scanning the frequency of the probe laser. By gating the free-space with the range resolution ΔR, the spectra within any slice of atmosphere can be obtained.

Fig. 2 shows a flow chart of data acquisition and processing.
Step 1: The backscattering profiles of the probe laser at different scanning steps and the reference laser at 190.652 THz are collected as photon counts $N(f_{pi}, R_j)$ and $N(f_0, R_j)$, respectively. Where i is the index of the frequency sampling steps and j represents the index of the range.
Step 2: The nonuniform optical gain of the EDFA $G(f_{pi})$ and the optical transmission $T_r(f_{pi})$ are calibrated as $N(f_{pi}, R_j)·G(f_{pi})·T_r(f_{pi})$ over the scanning span.

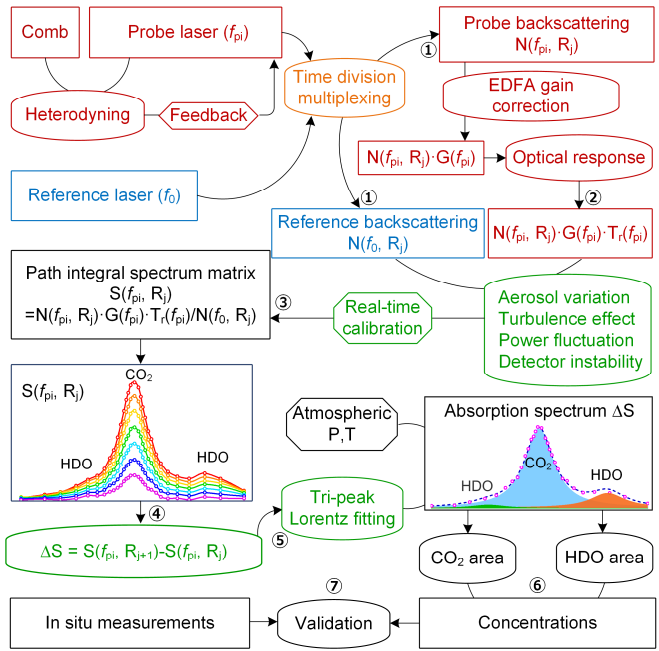

**Fig. 2.** Flow chart of data acquisition and processing.

Step 3: The aerosol variation, laser power fluctuation, instability and signal coupling efficiency instability due to turbulence are calibrated. By $N(f_{pi}, R_j) \cdot G(f_{pi}) \cdot T_r(f_{pi})/N(f_0, R_j)$, a matrix $S(f_{pi}, R_j)$ containing integrated absorption spectra over the optical path of $CO_2$ and HDO is obtained.

Step 4: The absorption spectrum $\Delta S$ at range $R = (R_j+R_{j+1})/2$ with range resolution $\Delta R = R_{j+1}-R_j$ is calculated by $S(f_{pi}, R_{j+1}) - S(f_{pi}, R_j)$.

Step 5: Triple-peak Lorentz nonlinear fitting is performed to separate the spectra of $CO_2$ and HDO. Several database-based apriori constraints, including the relative strength of two HDO lines, relative frequency offsets between peaks, and the FWHM calculated with in situ atmospheric temperature and pressure, are used in the fitting (see Supplement 1 for detailed information).

Step 6: Both $CO_2$ and HDO concentrations are determined by calculating the areas of the fitted spectra.

Step 7: Concentrations are compared to the results from in situ measurements for validation.

The raw backscattering signals of probe light $N(f_{pi}, R_j)$ and reference light $N(f_0, R_j)$ are shown in Figs. 3(a) and (b), respectively. The $CO_2$ absorption feature is clear at the center frequency of $N(f_{pi}, R_j)$ at approximately 190.667 THz. From step 1 to step 3, the spectra $S(f_{pi}, R_j)$ integrated over the optical depth of at different ranges can be obtained, as shown in Fig. 3(c). Then, in step 4, DFSS acquires the range-resolved spectra at different ranges. Fig. 3(d) illustrates one example of $\Delta S$, at 4 km with a range resolution of 60 m. After step 5, the $CO_2$ and HDO lines are separated by Lorentz nonlinear fitting, with standard errors of area estimation for $CO_2$ and HDO of 0.9% and 11%, respectively. The higher error for HDO is due to sparse frequency sampling and weak absorption of HDO over the frequency span.

A continuous field observation over 72 hours at the University of Science and Technology of China (USTC) (31.83°N, 117.25°E) is carried out. For easy comparison to in situ measurements, the laser beams are emitted horizontally. Figs. 4(a) and (b) show the horizontal range-time plots of the concentrations of $CO_2$ and HDO measured by DFSS. The range resolutions of both are 60 m, while the time resolutions for $CO_2$ and HDO are 10 and 30 min, respectively. The $CO_2$ and HDO concentrations are nearly the same throughout the whole detection range at any given time but fluctuate over the observation time. Many factors may contribute to these phenomena, such as dissipation caused by turbulence, atmospheric transport, human activities, industrial production and plant photosynthesis. Other instruments are employed to monitor the wind field and turbulence for further verifications.

Fig. 4(c) shows the carrier-to-noise ratio (CNR) simultaneously measured by coherent Doppler wind lidar (CDWL). The deposition of aerosol particles is inconspicuous in CNR, which represents rarely external transmissions of aerosol. Fig. 4(d) and e show the horizontal wind speed and horizontal wind direction measured by CDWL, providing additional information about the weather conditions. It is noteworthy that the horizontal wind speed in the near ground region is less than 5 m/s, and the wind direction is mainly easterly. Considering the lack of heavy industry east of the USTC, the relatively stable wind field shows that $CO_2$ and HDO are local and hardly affected by external transmission.

Fig. 4(f) shows the $CO_2$ concentrations at 2 km measured by DFSS and an in situ $CO_2$ analyzer (Thermo Scientific 410i) and the near-ground atmospheric refractive index structure constant $C_n^2$ measured by a large-aperture scintillometer (Kipp

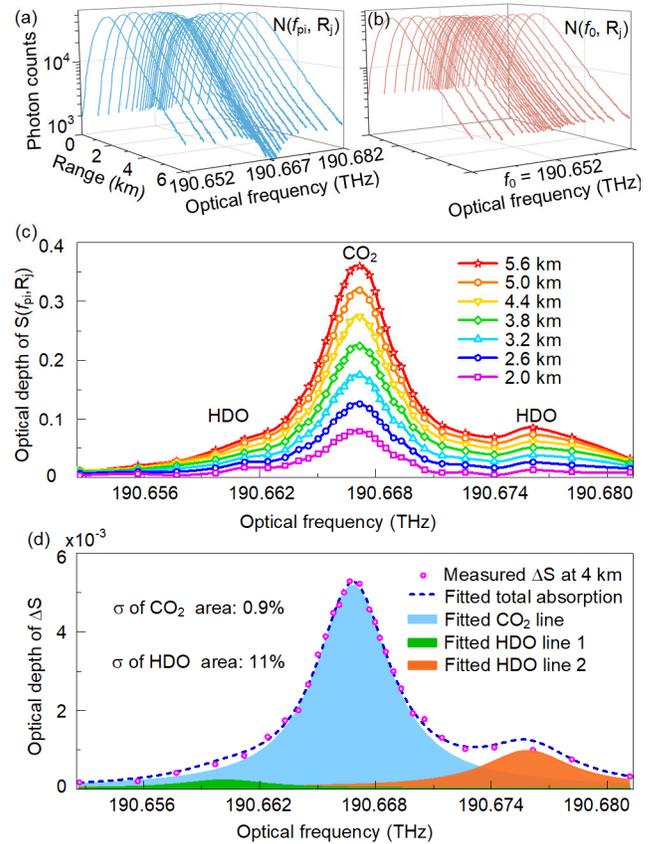

**Fig. 3.** Backscattering signals and spectra. (a) The probe signal, with 30 scanning frequencies, covers $CO_2$ and HDO absorption lines. (b) The reference signal without gas absorption. (c) The sum integrated optical depth line shapes of $CO_2$ and HDO at different ranges. (d) Lorentz fitting of the range-resolved spectrum; magenta dots are the measured $\Delta S$ values at 4 km with $\Delta R$ = 60 m.

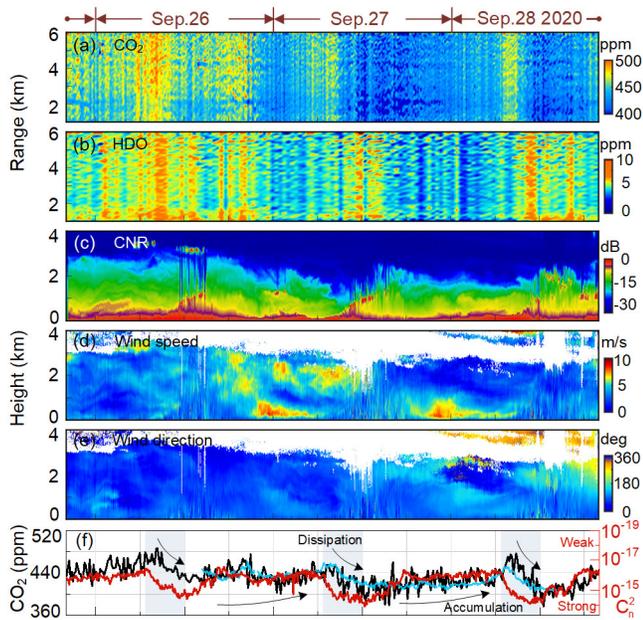

**Fig. 4.** Results of continuous observation. Range-time plots of the (a) $CO_2$. (b) HDO. (c) CNR. (d) Horizontal wind speed. (e) Horizontal wind direction. (f) $CO_2$ concentration comparison. The black and cyan lines represent the $CO_2$ concentration at 2 km measured by DFSS and the in situ $CO_2$ analyzer. The red line is $C_n^2$ measured by a scintillometer, with the y-coordinate reversed.

& Zonen LAS MKII). For easy analysis, the y-coordinate of $C_n^2$ is reversed. The observation of DFSS shows consistency with the gas analyzer results. In addition, during the continuous observation, the turbulence intensity gradually increases every morning from 8:00 to 12:00. Meanwhile, the $CO_2$ concentration dissipates rapidly. Then, the turbulence intensity decreases in the afternoon and remains weak during the night, while the $CO_2$ concentration accumulates gradually. On the one hand, the correlation between the turbulence intensity and the concentration of $CO_2$ is obvious. The $CO_2$ presents a diurnal variation throughout the observation with a daily periodicity of $C_n^2$. On the other hand, a time delay between the concentration change and turbulence intensity is measured, especially during the morning. Within the range of several kilometers (usually the resolution of satellite payload equipment), the concentrations of $CO_2$ and HDO change almost simultaneously. On the time scale, they are mainly affected by the atmospheric conditions of the boundary layer, especially turbulence.

In conclusion, a DFSS method has been proposed and demonstrated for remote sensing of multi-substance spectra at different locations over 6 km. The frequency accuracy of this DFSS method is guaranteed by employing heterodyne detection of a probe laser and an optical comb. To obtain precise spectra during scanning, a reference laser incorporating the TDM technique is alternated with the probe laser to reduce the influences of aerosol variation, laser power fluctuation, detector instability and telescope coupling efficiency change. In addition, an SNSPD with a large active area and low dark noise is employed to reduce the requirement of a high-power laser for long-range detection, making wide-optical-range DFSS possible with a low-power fiber laser. Future applications of DFSS include long-distance early warning of harmful substances, industrial pollution and explosive detection. Furthermore, human knowledge of substance evolution and chemical reactions in the atmosphere can be further improved.

**Acknowledgments.** This work was supported by the Anhui Initiative in Quantum Information Technologies.

**Disclosures.** The authors declare no conflicts of interest.

See Supplement 1 for supporting content.

# Supplement 1

## 1. THE PRINCIPLE OF COMB-REFERENCED FREQUENCY LOKING

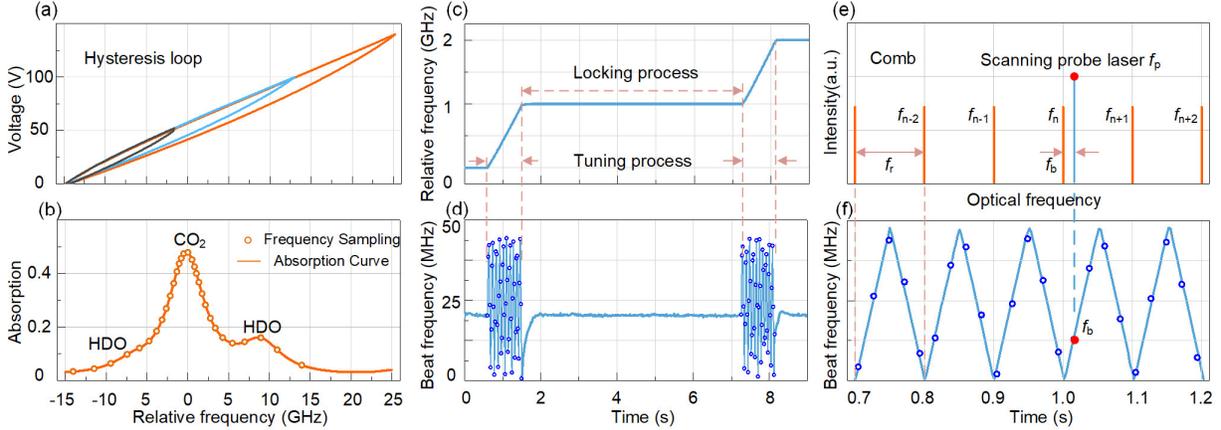

**Fig. S1.** Principle of comb-referenced frequency locking technique. (a) The "eye-pattern" loop curve shows the mapping between frequency and voltage of probe laser. The lines in different colors show the loops for controlling voltage in different ranges. (b) The absorption function of $CO_2$ and HDO from HITRAN. The hollow circles represent the frequencies sampled to scan the whole line shape, the numbers and interval of which can be varied to adapt different gases. (c) The optical frequency variation versus time. (d) The beat frequency variation versus time. The locking process is shown in the flat lines on either side, and the tuning process making the frequencies jumps over the preset interval is shown in blue hollow circles. (e) The probe laser $f_p$ beats with optical comb $f_n$, mapping optical frequency into radio frequency signals. (f) Detailed description of tuning process. As the probe laser tuned by controlling the PZT, the beat frequency $f_b$ varies as triangular.

Fig. S1 shows the principle of comb-referenced frequency locking technique. The ECDL is tuned by controlling its piezo-electric transducer (PZT) while the counter is used to monitor the beat frequency of the heterodyne detection. A typical "eye-pattern" of voltage versus relative optical frequency shown in Fig. S1(a) is measured when PZT is back and forth modulated. Different PZT tuning range is performed with repeatable "eye-patterns" due to hysteresis effect. The $CO_2$ absorption function centered at 190.667 THz contains two weak HDO lines according to HITRAN is shown in Fig. S1(b). By scanning the frequency with nonuniform step intervals (as shown in orange hollow circles), the scanning voltages can be determined through the "eye-pattern" measured in Fig. S1(a). Here we use 30 steps and the interval in Fig. S1(b) for better $CO_2$ scanning accuracy, making HDO an additional result. Significantly, follow-up more focused HDO spectrum can be realized by freely altering the scanning numbers and the interval of steps.

An electronic frequency locked loop is commonly used to stabilize the frequency of probe laser ($f_p$). An optical frequency comb with repetition rate $f_r$ = 100 MHz and carrier-envelope offset frequency $f_{ceo}$ = 20 MHz are phase locked to a rubidium clock. As shown in Fig. S1(c), during the tuning process, the optical beating between the probe laser and the comb is yielded and filtered by a 48 MHz low pass filter, making the recording beat frequency oscillates within $f_r$/2 presented in Fig. S1(d). As shown in Figs. S1(e) and (f), when $f_p$ increases, only beating signal $f_b$ of probe laser with the nearest tooth of the comb is recorded. During the locking process, the beating frequency yields a feedback signal, which tuning the optical frequency of probe laser by modulating the PZT voltage. An absolute frequency drift less than 1 MHz is realized.

## 2. TRIPE-PEAK LORENTZ NONLINEAR FITTING

The Lorentz model can be expressed as

$$\varphi(f) = \frac{2A}{\pi} \frac{\omega_L}{\omega_L^2 + 4(f - f_0)^2} \tag{S1}$$

In this expression $A = H\omega_L\pi/2$ is the area of Lorentz line shape with $H$ represents the height of the line center, $f_0$ is the center line of probe laser, and $\omega_L$ is the FWHM, which is determined by

$$\omega_L = 2P[\chi_m \gamma_m(T) + \chi_a \gamma_a(T)] \tag{S2}$$

Here $P$ is the total pressure of gases, $\chi$ is the ratio of the gas to the total atmospheric pressure, with subscript $m$ and $a$ mean target gas and air, respectively. $\gamma$ is the pressure broadening coefficient that is available from HITRAN. $T$ is the experiment temperature. Besides, $\chi$ depends on $T$ can also be described as

$$\gamma(T) = \gamma(T_0)(\frac{T_0}{T})^n \tag{S3}$$

$$\chi_m + \chi_a = 1 \tag{S4}$$

We suppose the target gas is negligible to air that $\chi_m$=0, $\chi_a$=1. Then, we can know the relative $\omega_L$ of $CO_2$ and HDO according to (2) and (3). Meanwhile, the fitted Lorentz area of absorption coefficient can be expressed as

$$A = \int_{line} \alpha(\nu) d\nu = \frac{P}{P_0} \frac{T_0}{T} n_L S(T) \tag{S5}$$

Here $n_L$ is called Loschmidt constant, $S(T)$ is the spectral intensity, in the case when $T_0$ = 273.15 K and $P_0$ = 100 kPa, $n_L$ = 2.6516467×$10^{25}$ molecule/m$^{-3}$, $S(T)$ = 1.77×$10^{-25}$ m/molecule for $CO_2$ and $S(T)$ = 2.32×$10^{-24}$ m/molecule for HDO. So that the pressures of $CO_2$ and HDO can be calculated and then the partial pressures representing concentrations are obtained.